\documentclass[sort,final]{aipproc}

\layoutstyle{6x9}

\usepackage{graphics,epsf,color}
\usepackage{amsmath,amssymb,natbib}

\def\Rset{\mathbb{R}}
\def\Dkl{D_{\mathrm{kl}}}
\def\Tr{\mbox{Tr}}
\def\Y{{\cal Y}}
\def\B{{\cal B}}
\def\L{{\cal L}}
\def\d{{\mbox{d}}}

\renewcommand{\vec}[1]{\mathbf{#1}}

\begin{document}

\title{Rigorous bounds for R\'enyi entropies of spherically symmetric potentials}

\classification{89.70.Cf,03.65.-w}
\keywords      {R\'enyi entropy, Shannon entropy, spherically symmetric potentials, variational upper bounds}

\author{Pablo S\'anchez-Moreno}{
address={Instituto Carlos I de F\'{\i}sica Te\'orica y Computacional, Univ. de Granada, 18071-Granada, Spain},
altaddress={Departamento de Matem\'atica Aplicada, Univ. de Granada, 18071-Granada, Spain},
email={pablos@ugr.es}
}

\author{Steeve Zozor}{
address={GIPSA-Lab, Domaine universitaire, 38402 St martin d'H\`{e}res, France},
email={steeve.zozor@gipsa-lab.inpg.fr}
}

\author{Jes\'us S. Dehesa}{
address={Instituto Carlos I de F\'{\i}sica Te\'orica y Computacional, Univ. de Granada, 18071-Granada, Spain},
altaddress={Departamento de F\'{\i}sica At\'omica, Molecular y Nuclear, Univ. de Granada, 18071-Granada, Spain},
email={dehesa@ugr.es}
}

\begin{abstract}
  The  R\'enyi and  Shannon entropies  are information-theoretic  measures which
  have  enabled to formulate  the position-momentum  uncertainty principle  in a
  much more adequate and stringent way than the (variance-based) Heisenberg-like
  relation.   Moreover,  they   are   closely  related   to  various   energetic
  density-functionals of  quantum systems.  Here we find  sharp upper  bounds to
  these quantities in terms of  the second order moment $\langle r^2\rangle$ for
  general spherically symmetric  potentials, which  substantially improve previous  results of
  this type,  by means of the  R\'enyi maximization procedure  with a covariance
  constraint due to Costa, Hero and Vignat \cite{CosHer03}. The contributions to
  these  bounds  coming  from the  radial  and  angular  parts of  the  physical
  wavefunctions  are   explicitly  given.
\end{abstract}

\maketitle

\section{Introduction}
\label{Introduction:sec}

The  Shannon and  R\'enyi  entropies of  a  normalized probability  distribution
$\rho(\vec{x})$,  $\vec{x}   =  (x_1   ,  \ldots  ,   x_d)  \in   \Rset^d$,  are
information-theoretic measures which quantify  the spread of $\rho(\vec{x})$ all
over the $d$-dimensional space in  different, but complementary, ways. They have
been  used  not  only  to  study the  statistical  properties  of  multifractals
\cite{JizAri04},  and to  derive entanglement  criteria for  continuous variable
systems \cite{saboia_arxiv10},  but also to  set up more adequate  and stringent
mathematical     formulations      \cite{BiaMyc75,Bia06,ZozPor08}     of     the
position-momentum    quantum-mechanical   uncertainty    principle    than   the
Heisenberg-like uncertainty relation.  Moreover,  although they are not physical
observables  because they  cannot  be  expressed as  expectation  values of  any
Hermitian operator  of a quantum system,  they are closely  related with various
macroscopic properties  of the system; particularly, with  energy functionals of
various types \cite{dehesa_10}.  These are  some physical reasons to motivate 
the search for sharp  upper bounds to these two entropies in  terms of the power
moment    of      order    $2$,     $\langle    r^2\rangle$,    where
$r^2=\|\vec{x}\|^2=\sum_{i=1}^d x_i^2$. This is because the power
moments are not only important by their own, but also they represent fundamental
and/or experimentally  accessible quantities of  the system (see e.g.  the recent
review \cite{dehesa_10}).

Extending  previous  three-dimensional  results  \cite{GadBen87,AngDeh92},  it  was
variationally  shown  \cite{CovTho91,Ang94}  that the  Shannon  entropy
defined as
\begin{equation}
S[\rho] = - \int_{\Rset^d} \rho(\vec{x}) \ln\rho(\vec{x}) \, \d\vec{x}
\end{equation}
is bounded from above by
\begin{equation}
S[\rho] \le \frac{d}{2} \ln \left( 2 \pi e \frac{\langle r^2 \rangle}{d} \right)
\label{Shannon_Bound_r2:eq}
\end{equation}
in  terms  of  $\langle  r^2\rangle$.  Moreover,  the  R\'enyi  entropy  defined
as \cite{Ren61,CovTho91,DemCov91}
\begin{equation}
H_\lambda[\rho] = \frac{1}{1-\lambda} \ln \int_{\Rset^d} \left[ \rho(\vec{x})
\right]^\lambda \, \d\vec{x}; \:\: \lambda > 0, \:\: \lambda \neq 1,
\label{Renyi_Definition:eq}
\end{equation}
has    shown   to    be    bounded   in    terms    of   $\langle    r^2\rangle$
\cite{dehesa_10,DehGal88,CosHer03,renyi_70} as follows:
\begin{equation}
H_\lambda[\rho] \leq \mathcal{B}_d(\lambda) + \frac{d}{2} \ln \left(
\frac{\langle r^2 \rangle}{d} \right),
\label{Renyi_Bound_r2:eq}
\end{equation}
with
\begin{equation}
\B_d(\lambda) = \left\{\begin{array}{ll}
\frac{d}{2} \log\left( \frac{\pi ((2+d) \lambda - d)}{\lambda-1}\right)
+ \log\left( \left(\frac{(2+d) \lambda - d}{2
\lambda}\right)^{\frac{1}{1-\lambda}} \frac{\Gamma \left( \frac{\lambda}{\lambda-1}
\right)}{\Gamma \left( \frac{(2+d) \lambda - d}{2 (\lambda-1)} \right)}
\right), & \lambda > 1\\[5mm]
\frac{d}{2} \log\left( \frac{\pi ((2+d) \lambda - d)}{1-\lambda}\right)
- \log\left( \left(\frac{(2+d) \lambda - d}{2
\lambda}\right)^\frac{\lambda}{1-\lambda} \frac{\Gamma \left( \frac{\lambda}{1-\lambda}
\right)}{\Gamma \left( \frac{(2+d) \lambda - d}{2 (1-\lambda)} \right)}
\right), & \lambda \in \left( \frac{d}{d+2} , 1
\right)\\[5mm]
\frac{d}{2}\log(2\pi e), & \lambda = 1
\end{array}\right.\label{Bound_MaxEnt:eq}
\end{equation}
Note   that  the  bounds   (\ref{Renyi_Bound_r2:eq})-(\ref{Bound_MaxEnt:eq})  to
R\'enyi entropies boil down  to the bound (\ref{Shannon_Bound_r2:eq}) to Shannon
entropy, in accordance to the known limit
$
 \lim_{\lambda\to 1} H_\lambda[\rho] =S[\rho].
$

On the other hand let us  highlight the fundamental relevance and utility of the
spherically symmetric potentials in the quantum-mechanical description of the natural systems. Indeed, the central-field model of  the atom is, together
with  the  Pauli  exclusion  principle,  the theoretical  basis  of  the  Aufbau
principle  of  the  Mendeleev  atomic  periodic  table.  Moreover,  the  spherically symmetric
potentials have been used as prototypes  for many other purposes and systems not
only  in   the  three-dimensional  world   but  also  in   non-relativistic  and
relativistic $d$-dimensional physics.

The    goal   of    this    paper    is   to    improve    the   upper    bounds
(\ref{Shannon_Bound_r2:eq})  and (\ref{Renyi_Bound_r2:eq})  to  the Shannon  and
R\'enyi  entropies in  terms of  $\langle r^2\rangle$  for quantum  systems with
spherically symmetric potentials.  Briefly, we use the following
three-step  methodology.   First,  we  separate  out  the   radial  and  angular
contributions  to the  physical entropies  by  making an  appropriate change  of
variable which  involves the covariance matrix  of the system. Then,  we use the
maximization procedure  of Costa et  al \cite{CosHer03} for the  R\'enyi entropy
under a covariant matrix constraint. Finally, we take into account the spherical
symmetry in the evaluation of the bound previously found.

The   structure  of   the   paper   is  the   following.    First, we  formulate the quantum-mechanical  $d$-dimensional problem
for  spherically symmetric  potentials,  indicating  the  probability density  of  the  system
$\rho(\vec{x})$, and  we  separate out  the
R\'enyi entropy of $\rho(\vec{x})$ into a radial and an angular part. Then, we obtain  an upper bound to the  radial R\'enyi entropy by
means   of  the   maximization  procedure   of  Costa   et  al   subject   to  a
covariance-matrix constraint,  and we calculate the angular  R\'enyi entropy in
terms of the generalized quantum  numbers $\{\mu\}$ of the system.  Finally,   some
conclusions and open problems are given.


\section{The quantum problem for $d$-dimensional spherically symmetric potentials}
\label{Quantum:sec}

The Schr\"odinger  equation of  a particle moving  in a  $d$-dimensional spherically symmetric
potential  $V_d(r)$,  i.e. that  only  depends on  the  distance  to the  origin
$r=\|\vec{x}\|$, $\vec{x} \in \Rset^d$, can be written as
\begin{equation}
\left[ -\frac12 \nabla_d^2 + V_d(r) \right] \Psi(\vec{x}) = E \Psi(\vec{x}),
\label{Schrodinger_Equation:eq}
\end{equation}
where  $\Psi(\vec{x})$  is  the  wavefunction  describing  a  quantum-mechanical
stationary  bound  state of  the  particle.  The  symbol $\vec{x}$  denotes  the
$d$-dimensional  position vector  having  the Cartesian  coordinates $\vec{x}  =
(x_1,x_2,\ldots,x_d)$       and       the       hyperspherical       coordinates
$(r,\theta_1,\theta_2,\ldots,\theta_{d-1})   \equiv   (r,\Omega_{d-1})$,   where
naturally $\|\vec{x}\|^2 = r^2 =  \sum_{i=1}^d x_i^2$; they are mutually related
by
\begin{equation}
\left\{\begin{array}{lll}
x_1&=&r\cos\theta_1\\
&\vdots&\\
x_k&=&r\sin\theta_1\ldots\sin\theta_{k-1}\cos\theta_k\\
&\vdots&\\
x_{d-1}&=&r\sin\theta_1\ldots\sin\theta_{d-2}\cos\theta_{d-1}\\
x_d&=&r\sin\theta_1\ldots\sin\theta_{d-2}\sin\theta_{d-1}\\
\end{array}\right.\label{Cartesian_Spherical:eq}
\end{equation}
where $r \in  [0 \: ; \: +\infty)$, $\theta_i  \in [0 \: ; \:  \pi), i < d-1$
and $\theta_{d-1} \in [0 \: ; \: 2 \pi)$.

It is known \cite{YanVan99,Ave02} that the wave function can be separated out
into a  radial, $R_{El}(r)$, and an  angular, $\Y_{\{\mu\}}(\Omega_{d-1})$, part
as
\begin{equation}
\Psi_{E,\{\mu\}}(\vec{x}) = R_{E,l}(r) \Y_{\{\mu\}}(\Omega_{d-1}).
\label{Wavefunction_Radial_Angular:eq}
\end{equation}

Note that the  angular part, which is common to any  spherically symmetric potential, is given
by       the       hyperspherical       harmonics       \cite{YanVan99,Ave02}
$\Y_{\{\mu\}}(\Omega_{d-1})$, which satisfy the eigenvalue equation
\[
\Lambda^2_{d-1}   \Y_{l,\{\mu\}}(\Omega_{d-1})   =  l   (   l   +   d  -   2   )
\Y_{l,\{\mu\}}(\Omega_{d-1}),
\]
associated to the generalized angular momentum operator given by
\[
\Lambda^2_{d-1}  = -  \sum_{i=1}^{d-1} \frac{  (\sin\theta_i)^{i+1-d}  }{ \left(
    \prod_{j=1}^{i-1} \sin\theta_j \right)^2 } \frac{\partial}{\partial\theta_i}
\left[ (\sin\theta_i)^{d-i-1} \frac{\partial}{\partial\theta_i} \right].
\]
The  quantum  angular  numbers  $(l\equiv\mu_1,\mu_2,\ldots,\mu_{d-1}\equiv  m)$
satisfy the chain of inequalities
$l\equiv\mu_1\ge\mu_2\ge\cdots\ge \mu_{d-2}\ge|\mu_{d-1}|\equiv|m|$.

These      mathematical      functions      can      be      expressed      (see
e.g. \cite{YanVan99,Ave02,DehLop10}) as
\begin{equation}
\Y_{\{\mu\}}(\Omega_{d-1})=\frac{e^{im\theta_{d-1}}}{\sqrt{2\pi}}
\prod_{j=1}^{d-2}\frac{1}{\sqrt{Z(\lambda_j,n_j)}}
\: C_{n_j}^{\lambda_j}(\cos \theta_j) \: (\sin \theta_j)^{\mu_{j+1}},
\label{Hyperharmonics:eq}
\end{equation}
with
\begin{equation}
n_j=\mu_j-\mu_{j+1},\quad \lambda_j = \frac{d-1-j}{2} + \mu_{j+1} .
\label{nj_lambdaj:eq}
\end{equation}
and the normalization constant  \cite[eq. 8.939-8]{GraRyz80}
\begin{equation}
Z(\lambda,n) = \int_0^\pi \left( C_n^\lambda(\cos \theta) ( \sin \theta)^\lambda
\right)^2 \d\theta =
\frac{\pi \, 2^{1-2\lambda} \Gamma(n+2\lambda)}{(\lambda+n) n!
(\Gamma(\lambda))^2},
\label{Normalization_Z:eq}
\end{equation}
where $C_n^\lambda(x)$ are the Gegenbauer polynomials.
Taking      the      Ansatz     (\ref{Wavefunction_Radial_Angular:eq})      into
(\ref{Schrodinger_Equation:eq}),  one obtains that  the radial  part $R_{El}(r)$
fulfils the second order differential equation
\[
\left[ -\frac12 \frac{\d^2}{\d r^2} - \frac{d-1}{2 r} \frac{\d}{\d r} + \frac{ l
    (l+d-2) }{ 2 r^2} + V_d(r) \right] R_{E,l}(r) = E \, R_{E,l}(r),
\]
which only  depends on the  energy $E$, the  dimensionality $d$ and  the angular
quantum number $l=\mu_1$.

Then, the quantum-mechanical probability density is given by
\begin{equation}
\rho_{E,\{\mu\}}(\vec{x})=|\Psi_{E,\{\mu\}}(\vec{x})|^2=|R_{E,l}(r)|^2
|\Y_{\{\mu\}}(\Omega_{d-1})|^2.
\label{Density_Spherical:eq}
\end{equation}

Let  us  note  that the  second-order  moment $\langle  r^2\rangle$ of  the
$d$-dimensional density $\rho_{E,\{\mu\}}(\vec{x})$ has the expression
\[
\langle  r^2 \rangle  =  \int_{\Rset^d} \|\vec{x}\|^2  \rho_{E,\{\mu\}}(\vec{x})
\,\d\vec{x} = \int_0^{+\infty} r^2 |R_{El}(r)|^2 r^{d-1} \, \d r,
\]
which  only depends  on the  radial wave  function $R_{El}(r)$  of  the particle
because the angular contribution is equal  to the unity due to the normalization
of    the    hyperspherical    harmonics.    Taking    into    account    bounds
(\ref{Shannon_Bound_r2:eq})  and (\ref{Renyi_Bound_r2:eq})  to  the Shannon  and
R\'enyi entropies, it  seems natural to think that to improve  such bounds we have
to use a  variational method with constraints which involve  not only the radial
part but also the angular part of the wave function. A way to do that is to consider the whole
covariance matrix $R_x=\langle \vec{x}\vec{x}^t\rangle$,
which has  the components  $\langle x_i x_j\rangle$,  not only the  second order
moment.


\section{R\'enyi entropy and covariance matrix}
\label{RenyiCov:sec}

Let us  now initiate  the determination of  the upper  bound to the  Shannon and
R\'enyi entropies  of the probability  density $\rho_{E,\{\mu\}}(\vec{x})$ given
by eq.  (\ref{Density_Spherical:eq}), by separating  out the radial  and angular
contributions  to  these  quantities.  To  do  that  we  use,  for  mathematical
convenience, a more appropriate, statistical notation.

Consider  a column  vector  $\vec{x} \in  \Rset^d$,  a $d$-dimensional  position
described by the wave function $\Psi(\vec{x})$, and interpret such a vector as a
random   vector    of   probability   density    function   $\rho_x(\vec{x})   =
|\Psi(\vec{x})|^2$. Without loss of generality, assume that the vector $\vec{x}$
is centered,  i.e. $\langle \vec{x} \rangle  = 0$.  In  general, the statistical
second order  moment (or  second order ensemble  average) $\langle  r^2 \rangle$
does not describe  entirely the second order statistics  of vector $\vec{x}$ and
it is  common to  consider all  the second order  statistics via  the covariance
matrix  $R_x  = \left\langle  \vec{x}  \vec{x}^t  \right\rangle$, of  components
$\langle x_i x_j \rangle$. Note that these two statistical quantities are linked
by $\langle r^2  \rangle = \Tr(R_x)$, where $\Tr$ denotes the  trace of a matrix
(i.e.  the sum of its diagonal components). Let us now consider the ``modified''
position vector $\vec{y} = R_x^{-1/2}  \vec{x}$, where $R_x^{1/2}$ is the unique
symmetric  positive  definite  matrix  that  is  the  ``square-root''  of  $R_x$
\cite[th.   7.2.6]{HorJoh85}.   From   the  properties  of  covariance  matrices
\cite{Fel71}, this is equivalent to  consider a stretched and rotated version of
the position vector $\vec{x}$. Clearly the covariance matrix of $\vec{y}$ is the
identity matrix $I_d$,  meaning that the transformation of  $\vec{x}$ is so that
the position $\vec{y}$ is isotropic in terms of the second order statistics.

Then,   the   R\'enyi   entropy   of   $\vec{x}$   (or   $\rho_x$),   given   by
eq.  (\ref{Renyi_Definition:eq}),  and  that  of  $\vec{y}$  (or  $\rho_y$)  are
mutually related by
\begin{equation}
H_\lambda[\rho_x] = H_\lambda[\rho_y] + \frac12 \log |R_x|
\label{Entropy_logR:eq}
\end{equation}
(see e.g.   \cite{Ren61,CovTho91,DemCov91}).  To obtain this  well known result,
one uses the change  of variable $\vec{x} = R_x^{1/2} \vec{y}$
in  $H_\lambda(\rho_x)$,  and one  realizes  that $\rho_y(\vec{y})  =|R_x|^{1/2}
\rho_x(R_x^{1/2}    \vec{y})$.    Notice    that    since   $H_1[\rho]=S[\rho]$,
eq. (\ref{Entropy_logR:eq}) is also valid  for the Shannon entropies of $\vec{x}$
and $\vec{y}$ (see  e.g.  \cite[eq.  (9.67)]{CovTho91} in the  Shannon context or
\cite[eq.  (22)]{DemCov91}).   Let us also  keep in mind  that the trace  of the
covariance matrix  is the expectation  value of the  square norm of  the vector;
then,    $\vec{y}$    verifies    that    its   expectation    value    $\langle
\|\vec{y}\|^2\rangle = d$.

Now, we  can extract  the ``variance''  term $\langle r^2  \rangle$, where  $r =
\|\vec{x}\|$, from the covariance matrix remembering that $\langle r^2 \rangle =
\Tr(R_x)$, and writing
\begin{equation}
R_x = \frac{\langle r^2 \rangle}{d} \: d \: C_x \hspace{5mm} \mbox{ where }
\hspace{5mm} C_x = \frac{R_x}{\Tr(R_x)}.
\label{CxRx:eq}
\end{equation}
Matrix $C_x$  contains only the  correlation structure of  $\vec{x}$, regardless
its  strength. $C_{x,i,i}$ represents  the relative  strength of  component $x_i$
relatively to the total power. Obviously, as $|R_x|=(\langle r^2 \rangle/d)^d \:
d^d \: |C_x|$, we have from (\ref{Entropy_logR:eq}) that
\begin{equation}
H_\lambda(\rho_x) =  H_\lambda(\rho_y) + \frac{d}{2} \log \frac{\langle r^2 \rangle}{d}
+\L(\Omega_{d-1}),
\label{H_Rad_Corr:eq}
\end{equation}
where the symbol
\begin{equation}
\L(\Omega_{d-1}) = \frac12 \log  |C_x| + \frac{d}{2} \log d.
\label{Loss_Determinant:eq}
\end{equation}
represents  the  contribution to  the  entropy  coming  from the  hyperspherical
harmonics.  Indeed,  for a  spherically symmetric density  $\rho_x(\vec{x})$ the
covariance  matrix is  $R_x =  \frac{\langle r^2  \rangle}{d} I_d$  (so,  $C_x =
\frac{1}{d}  \, I_d$)  and  then  its R\'enyi  entropy  is $H_\lambda[\rho_y]  +
\frac{d}{2} \log \frac{\langle r^2\rangle}{d}$.

Let us denote by $\lambda_{x,i} \ge 0$ the eigenvalues of $C_x$. We have $\sum_i
\lambda_{x,i} = \Tr(C_x) = 1$ and thus they can be viewed as probabilities.  Let
us  denote by  $\lambda_x  =  \{\lambda_{x,1} ,  \ldots  , \lambda_{x,d}\}$  the
discrete  density of  eigenvalues and  by $u_d  = \{1/d  , \ldots  ,  1/d\}$ the
uniform density.  Hence, we obtain that
\begin{equation}
\L(\Omega_{d-1}) = - \frac{d}{2} \sum_{i=1}^d \frac{1}{d} \log
\left(\frac{1/d}{\lambda_{x,i}} \right) = - \frac{d}{2} \Dkl( u_d\| \lambda_x)\le 0.
\label{Loss_Eigenvalues:eq}
\end{equation}
The symbol  $\Dkl(v\|w)$ denotes the Kullback-Leibler (KL,  in short) divergence
between the distributions $v$ and $w$ as defined by $\Dkl(v\|w)=\sum_{i=1}^d v_i\log\left(v_i/w_i\right)$,
which is always  positive unless the two distributions are  equal, in which case
it    vanishes   \cite[th.     2.6.3]{CovTho91}.    Then,    the   KL-divergence
$\Dkl(u_d\|\lambda_x)$ between  the uniform distribution $u_d$  and the counting
density  of eigenvalues  of $C_x$, which  is controlled  by the  hyperspherical
harmonics only,  quantifies the loss of entropy,  $\L(\Omega_{d-1})$. Note now that the  loss of entropy vanishes if  and only if
the  eigenvalues of  $C_x$ are  uniformly distributed,  what implies  that  $\Dkl(u_d  \| \lambda_x)  =0$, and  thus that  matrix $C_x$  is
diagonal and equals to $\frac{1}{d}  I_d$.  Reciprocally, for $C_x = \frac{1}{d}
I_d$, one  has $\L(\Omega_{d-1})  = 0$. 

Therefore,   according  to  eq.   (\ref{H_Rad_Corr:eq}),  the   R\'enyi  entropy
$H_\lambda[\rho_x]$  of the  quantum  probability density  $\rho(\vec{x})$ of  a
particle in a  spherically symmetric potential can be separated out into  two parts: one which
contains the contribution of the  radial wavefunction of the system, and another
one ($\L(\Omega_{d-1})$) which only depends on the angular wavefunction, that is
on the hyperspherical harmonics. The radial R\'enyi entropy cannot be calculated
unless we  know the specific analytical  form of the spherically symmetric  potential, but the
angular     R\'enyi    entropy    can     be    explicitly     determined    via
eq.  (\ref{Loss_Determinant:eq}).  These  two  issues  are tackled  in  the  next
Section.


\section{Upper bounds to the R\'enyi entropy}
\label{UpperBound:sec}

In this Section we bound from above the R\'enyi entropy of the isotropic density
$\rho_y(\vec{y})$,  $H_\lambda[\rho_y]$,  and  then  we determine  the  loss  of
entropy  $\L(\Omega_{d-1})$  due to  the  hyperspherical  entropy; so,  obtaining
together with eq.  (\ref{H_Rad_Corr:eq}) a sharp upper bound  to the total R\'enyi
and Shannon entropies of a particle moving in a spherically symmetric potential.

Let  us begin  with  the  use of  the  extremization procedure  of  Costa et  al
\cite{CosHer03} to bound the  R\'enyi entropy $H_\lambda[\rho_y]$ subject to the
covariance-matrix constraint $R_y =  I_d$. It straightforwardly yields the upper
bound
\begin{equation}
H_\lambda(\rho_y) \le \B_d(\lambda)
\label{MaxEnt:eq}
\end{equation}
where $\B_d(\lambda)$ is given by (\ref{Bound_MaxEnt:eq}).

Now, let  us calculate  the loss of  entropy $\L(\Omega_{d-1})$.   Starting from
eq. (\ref{Loss_Determinant:eq})  it is easily shown that  all the  non-diagonal  matrix  elements of  $C_x$  for spherically symmetric  potentials
vanish, what implies that the matrix $C_x$ is diagonal, in which case
\begin{equation}
\L(\Omega_{d-1})  = \frac{1}{2} \sum_{i=1}^d \log C_{x,i,i} + \frac{d}{2} \log d
\label{Entropy_Loss_Diagonal_C:eq}
\end{equation}
Then,  keeping  in mind  eq.  (\ref{Cartesian_Spherical:eq})  we  have that  the
diagonal elements $C_{x,i,i}$ of the matrix $C_x$ are given by
\begin{equation}
C_{x,i,i} = \frac{\langle x_i x_i\rangle}{\langle r^2\rangle} = \left(
\prod_{k=1}^{i-1} \langle \sin^2\theta_k \rangle \right) \langle \cos^2\theta_i
\rangle,
\label{Cxii:eq}
\end{equation}
where the trigonometric  expectation values are defined in  terms of the partial
hyperspherical harmonics, and where the
non existing angle  $\theta_d$ has to be removed from this  expression (or to be
chosen as $\theta_d = 0$ by convention).

Since  $\langle  \sin^2\theta_k\rangle=1-\langle\cos^2\theta_k\rangle$, we  only
need to  evaluate $\langle \cos^2\theta_k\rangle$ for $k<d-1$.  The latter value
can be obtained as follows
\[
\langle \cos^2 \theta_k \rangle =  \frac{1}{Z(\lambda_k,n_k)} \int_0^\pi
\left( \cos\theta \, C_{n_k}^{\lambda_k}(\cos \theta) \, \sin^{\lambda_k}
\theta\right)^2 \d\theta =  \frac{n_k^2+2 \lambda_k n_k + \lambda_k-1}{2 \,
(n_k+\lambda_k+1)(n_k+\lambda_k-1)}\\
\]

Replacing  $n_k$  and  $\lambda_k$  by their  values  (\ref{nj_lambdaj:eq})  one
obtains
\begin{equation}
\langle \cos^2 \theta_k \rangle = \frac{2 \mu_k (\mu_k+d-k-1) - 2 \mu_{k+1}
(\mu_{k+1}+d-k-2) +d-k-3}{4 \mu_k (\mu_k+d-k-1)+(d-k+1)(d-k-3)}.
\label{Mean_cos2:eq}
\end{equation}
For $k=d-1$,  and with the convention $\mu_d  = 0$, one has  $\lambda_{d-1} = 0$
and a  direct computation  shows that (\ref{Mean_cos2:eq})  holds also for  $k =
d-1$, i.e. $\langle \cos^2 \theta_{d-1} \rangle = \frac{1}{2}$.

Then,  from eqs.  (\ref{Loss_Determinant:eq}), (\ref{Entropy_Loss_Diagonal_C:eq})
the angular R\'enyi entropy or loss of entropy turns out to have the expression
\begin{equation}
\L(\Omega_{d-1}) = \frac12 \sum_{k=1}^{d-2} \left( (d-k) \log \langle \sin^2
\theta_k \rangle + \log \langle \cos^2 \theta_k \rangle\right) - \log 2 +
\frac{d}{2} \log d,
\label{Loss_Entropy_Central:eq}
\end{equation}
where   the    trigonometric   expectation    values   can   be    obtained   by
eq. (\ref{Mean_cos2:eq}).

Then, taking into account  eq. (\ref{H_Rad_Corr:eq}) and ineq. (\ref{MaxEnt:eq})
one finally has the sharp bound
\begin{equation}
H_\lambda[\rho_x] \le \B_d(\lambda) + \frac{d}{2} \log \frac{\langle
r^2\rangle}{d} + \L(\Omega_{d-1}),
\label{Renyi_General_Bound:eq}
\end{equation}
for  $d$-dimensional  single-particle  systems  in a  spherically symmetric  potential,  where
$\L(\Omega_{d-1})$    is   obtained    from   eqs.    (\ref{Mean_cos2:eq})   and
(\ref{Loss_Entropy_Central:eq}).

\section{Conclusions and open problems}
\label{Conclusion:sec}

In this work we substantially improve  the sharp upper bounds to the Shannon and
R\'enyi entropies of the stationary quantum states of single-particle systems in
a spherically symmetric potential, previously found in terms of the expectation value $\langle
r^2\rangle$. This is done by taking into  account the loss of entropy due to the
explicit consideration  of the angular  part of the  corresponding wavefunctions
(i.e., the hyperspherical  harmonics) into these physical entropies  by means of
the  covariance matrix  of the  system. Briefly,  we have  decomposed  the R\'enyi
entropy into two parts: one depending on the radial wavefunction and another one
on the angular wavefunction. Then, the radial R\'enyi entropy is upper bounded in
terms  of  $\langle  r^2\rangle$  and  the angular  part  is  explicitly
calculated making profit of the the  fact that the covariance matrix is diagonal
for spherically symmetric  potentials. 

A natural and useful extension of this work is to use radial expectation values of order other than 2 as constraints, which would require the ideas and methodology of Refs. \cite{dehesa_pra89,CosHer03}; or considering $q$-variances instead of
classical variance  by changing  the pdf $\rho$  by $\rho^q$ in  the statistical
average. Another way of extending the work could be to look to the optimum structure of the correlation matrix that minimizes the bound.


\begin{theacknowledgments}
PSM and JSD are very grateful to Junta de Andaluc\'{\i}a for the grants FQM-2445
and  FQM-4643,  and the  Ministerio  de Ciencia  e  Innovaci\'on  for the  grant
FIS2008-02380. PSM and JSD belong to the research group FQM-207.
\end{theacknowledgments}

\bibliographystyle{aipproc}
\bibliography{preprint_entropy_central}

\end{document}